\documentstyle[amssymb,preprint,aps]{revtex}

\begin{document}
\title{{\LARGE ENERGY OF KERR-NEWMAN BLACK-HOLES AND GRAVITOMAGNETISM}}
\author{Marcelo Samuel Berman}
\address{Editora Albert Einstein Ltda.\\
R. Candido Hartman,575 \#17 \ Ed. Renoir-Champagnat\\
80730-440 Curitiba PR BRAZIL\\
ABSTRACT\\
New formulae are obtained for the energy of K.N. b.h.'s that point out a\\
gravitomagnetic energy effect. The results are valid for slowly or rapidly\\
rotating black-holes. The expression of the energy density of Kerr-Newman\\
back-holes in the slow rotation case, is obtained afterwards, and shown to\\
be essentially positive.Subsequently,we show how to attain a ''repulsive''\\
gravitation (antigravitation) state identified with negative energy\\
distribution contents in a limited region of space, without violating the\\
Positive Energy Theorem.}
\maketitle

\begin{center}
\bigskip

\bigskip \bigskip

\bigskip

\bigskip

\bigskip

\bigskip \bigskip

\bigskip

\bigskip

\bigskip

\bigskip

\bigskip

\bigskip

\bigskip

\bigskip

\bigskip

\bigskip

\bigskip

\bigskip

\bigskip

\bigskip

\bigskip

\bigskip

\bigskip

\bigskip

\bigskip

\bigskip

\bigskip

\bigskip

\bigskip

\bigskip

\bigskip

\bigskip

\bigskip

\bigskip

\bigskip

\bigskip

{\LARGE ENERGY OF KERR-NEWMAN BLACK-HOLES AND GRAVITOMAGNETISM}

\bigskip

Marcelo Samuel Berman

\bigskip

\bigskip
\end{center}

There are pretty standard pseudotensor calculations presented in the book \
by Adler et al.$^{(1)}$.The most general black hole is characterized by mass
M, electric charge Q and rotational parameter ''a'' and is given by
Kerr-Newman's metric, where in quasi-Cartesian \ form, is given by:

\bigskip

ds$^{2}=dt^{2}-dx^{2}-dy^{2}-dz^{2}-\frac{2\left[ M-\frac{Q^{2}}{2r_{0}}%
\right] r_{0}^{3}}{r_{0}^{4}+a^{2}z^{2}}\cdot F^{2}$ \ \ \ \ \ \ \ \ \ \ \ \
\ \ \ \ \ \ \ \ \ \ \ \ \ \ \ \ \ \ \ \ \ \ \ \ \ \ \ \ \ \ \ \ \ (1)

\bigskip

F = dt + $\frac{Z}{r_{0}}dz+\frac{r_{0}}{\left( r_{0}^{2}+a^{2}\right) }%
\left( xdx+ydy\right) +\frac{a\left( xdy-ydx\right) }{a^{2}+r_{0}^{2}}$ \ \
\ \ \ \ \ \ \ \ \ \ \ \ \ \ \ \ \ \ \ \ \ \ \ \ \ \ \ \ \ \ \ \ \ \ \ \ (2)

\bigskip

r$_{0}^{4}-\left( r^{2}-a^{2}\right) r_{0}^{2}-a^{2}z^{2}=0$ \ \ \ \ \ \ \ \
\ \ \ \ \ \ \ \ \ \ \ \ \ \ \ \ \ \ \ \ \ \ \ \ \ \ \ \ \ \ \ \ \ \ \ \ \ \
\ \ \ \ \ \ \ \ \ \ \ \ \ \ \ \ \ \ \ \ (3)

\bigskip

and

\bigskip

r$^{2}\equiv x^{2}+y^{2}+z^{2}$ \ \ \ \ \ \ \ \ \ \ \ \ \ \ \ \ \ \ \ \ \ \
\ \ \ \ \ \ \ \ \ \ \ \ \ \ \ \ \ \ \ \ \ \ \ \ \ \ \ \ \ \ \ \ \ \ \ \ \ \
\ \ \ \ \ \ \ \ \ \ \ \ \ \ \ \ \ \ \ \ \ \ \ (4)

\bigskip

The energy-momentum quadrivector P$_{\mu }$, the energy-tensor T$_{\mu }^{V}$
and the energy-momentum pseudo-tensor of the gravitational field \ t$_{\mu
}^{V}$ obey the following relations:

\bigskip

P$_{\mu }=%
\displaystyle\int %
\limits_{t}\sqrt{-g}\left[ T_{\mu }^{0}+t_{\mu }^{0}\right] d^{3}x=$ \
constants. \ \ \ \ \ \ \ \ \ \ \ \ \ \ \ \ \ \ \ \ \ \ \ \ \ \ \ \ \ \ \ \ \
\ \ \ \ \ \ \ \ \ \ \ \ \ \ \ \ (5)

\bigskip

$\sqrt{-g}t_{\mu }^{V}=\frac{1}{2\complement }\left[ Ug_{\mu }^{\nu }-\frac{%
\partial U}{\partial g_{|\nu }^{\pi \beta }}g_{|\mu }^{\pi \beta }\right] $
\ \ \ \ \ \ \ \ \ \ \ \ \ \ \ \ \ \ \ \ \ \ \ \ \ \ \ \ \ \ \ \ \ \ \ \ \ \
\ \ \ \ \ \ \ \ \ \ \ \ \ \ \ \ \ \ \ \ \ \ \ \ \ (6)

\bigskip

U = $\sqrt{-g}g^{\rho \sigma }\left[ \left\{ 
\begin{array}{c}
\alpha  \\ 
\sigma \varrho 
\end{array}
\right\} \left\{ 
\begin{array}{c}
\beta  \\ 
\alpha \beta 
\end{array}
\right\} -\left\{ 
\begin{array}{c}
\alpha  \\ 
\beta \varrho 
\end{array}
\right\} \left\{ 
\begin{array}{c}
\beta  \\ 
\alpha \sigma 
\end{array}
\right\} \right] $ \ \ \ \ \ \ \ \ \ \ \ \ \ \ \ \ \ \ \ \ \ \ \ \ \ \ \ \ \
\ \ \ \ \ \ \ (7)

\bigskip

$\complement =-\frac{8\pi G}{c^{2}}$ \ \ \ \ \ \ \ \ \ \ \ \ \ \ \ \ \ \ \ \
\ \ \ \ \ \ \ \ \ \ \ \ \ \ \ \ \ \ \ \ \ \ \ \ \ \ \ \ \ \ \ \ \ \ \ \ \ \
\ \ \ \ \ \ \ \ \ \ \ \ \ \ \ \ \ \ \ \ \ \ \ \ \ \ \ \ \ \ \ \ (8)

\bigskip

After a lengthy calculation, we find: (in G=c=1 units)

\bigskip

P$_{0}=M-\left[ \frac{Q^{2}+M^{2}}{4\varrho }\right] \left[ 1+\frac{\left(
a^{2}+\varrho ^{2}\right) }{a\varrho }arctgh\left( \frac{a}{\varrho }\right) %
\right] $ \ \ \ \ \ \ \ \ \ \ \ \ \ \ \ \ \ \ \ \ \ \ \ \ \ \ \ \ \ \ \ \ \
\ \ \ \ \ \ \ \ \ \ \ \ (9)

\bigskip

P$_{1}=P_{2}=P_{3}=0$ \ \ \ \ \ \ \ \ \ \ \ \ \ \ \ \ \ \ \ \ \ \ \ \ \ \ \
\ \ \ \ \ \ \ \ \ \ \ \ \ \ \ \ \ \ \ \ \ \ \ \ \ \ \ \ \ \ \ \ \ \ \ \ \ \
\ \ \ \ \ \ \ \ \ \ \ \ \ \ \ (10)

\bigskip

By considering an expansion of the arctgh($\frac{a}{\varrho }$) function, in
terms of increasing powers of the parameter ''a'', and by neglecting terms a$%
^{3}\simeq a^{4}\simeq ....\simeq 0$, we find the energy of a slowly
rotating Kerr-Newman black-hole,

\bigskip

E $\simeq M-\left[ \frac{Q^{2}+M^{2}}{R}\right] \left[ \frac{a^{2}}{3R^{2}}+%
\frac{1}{2}\right] $ \ \ \ \ \ \ \ \ \ \ \ \ \ \ \ \ \ \ \ \ \ \ \ \ \ \ \ \
\ \ \ \ \ \ \ \ \ \ \ \ \ \ \ \ \ \ \ \ \ \ \ \ \ \ \ \ \ \ \ \ \ \ (11)

\bigskip

where $\varrho \rightarrow R$; this can be seen because the defining
equation for $\varrho $\ \ is:

\bigskip

$\frac{x^{2}+y^{2}}{\varrho ^{2}+a^{2}}+\frac{z^{2}}{\varrho ^{2}}=1$ \ \ \
\ \ \ \ \ \ \ \ \ \ and if a $\rightarrow $\ 0, $\varrho \rightarrow R$.

\bigskip

\bigskip

We can interpret the terms $\frac{Q^{2}a^{2}}{3R^{3}}$ and $\frac{M^{2}a^{2}%
}{3R^{3}}$ as the magnetic and gravitomagnetic energies caused by rotation.
Virbhadra$^{(2)}$ noticed the first of these effects in the year 1990, but
since then it seems that he failed to recognize the existence of the
gravitomagnetic energy due to M, on an equal footing.

\bigskip

Furthermore, we can, from \ relation (11), find the energy density \
associated with the black-hole:

\bigskip $\mu =\frac{1}{4\pi R^{2}}\frac{dE}{dR}=\frac{1}{4\pi R^{4}}\left[
Q^{2}+M^{2}\right] \left[ \frac{a^{2}}{R^{2}}+\frac{1}{2}\right] $\ \ \ \ \
\ \ \ \ \ \ \ \ \ \ \ \ \ \ \ \ \ \ \ \ \ \ \ \ \ \ \ \ \ \ \ \ \ \ \ \ \ \
\ \ (12)

\bigskip

As expected, the energy density is essentially positive, so it obeys the
weak energy condition.

\bigskip

\bigskip We shall \ \ now show how to attain a ''repulsive'' gravitation (or
antigravitation) state identified with negative energy contents in a limited
region of space, without violation of the positivity of energy.

The total energy E of an isolated system should be positive in General
Relativity, in accordance with the ''Positive Energy Theorem'', of Schoen
and Yau, Choquet-Bruhat, Deser, Teitelboim, Witten, York, etc$^{(3)(12)}$,
as cited by Ciufolini and Wheeler$^{(13)}$. The exception is Minkowski
space, where E = 0. ( See also Weinberg$^{(14)}$). The condition usually
stated for these theorems is that the dominance of energy condition be
satisfied, which entails the weak energy condition $\mu \geqslant 0$ \ \ \ ,
where \ $\mu $\ \ \ stands for the energy density$^{(13)}$, while the
Einstein field equations are taken as valid.

Even in popular books, like Hawking's best-seller$^{(15)}$, it is stated
that (total) energy in General Relativity is positive. The rule is broken by
the energy of the Universe, E = 0, as stated in Hawking's book$^{(15)}$,
because zero is not positive, and the Universe's metric is not Minkowskian (
see Berman$^{(16)}$).

Ciufolini and Wheeler$^{(13)}$ also comment that an electric charge circling
in orbit creates magnetism and, likewise, a spinning mass creates
gravitomagnetism; the gravitomagnetic effect due to the slow rotation of the
Earth is measureable, by an orbiting gyroscope.( Lageos III; Gravity Probe
B, etc. ). However, we are far from a spinning black-hole$^{(13)}$.

When obtaining the central mass field, or Kerr metric, for a spinning
massive object, one has to measure the energy contents of the distribution,
encircled by a radial distance R, and we have shown that, for
non-relativistic (slow) rotations, the energy contents is given by (within a
radial distance R):

\bigskip

\bigskip E=Mc$^{2}-\frac{GM^{2}}{2R}-\frac{a^{2}M^{2}}{3R^{3}}$ \ \ \ \ \ \
\ \ \ \ \ \ \ \ \ \ \ \ \ \ \ \ \ \ \ \ \ \ \ \ \ \ \ \ \ \ \ \ \ \ \ \ \ \
\ \ \ \ \ \ \ \ \ \ \ \ \ \ \ \ \ \ \ \ \ \ \ \ \ \ (13).

\bigskip

where M, c G, R and ''a'', stand respectivelly for central mass, speed of
light in vacuum, gravitational constant, radial distance, and angular
momentum per unit mass, and \ Q=0.

The first rhs term above is the inertial energy, as known from Special
Relativity; the second, is the relativistic equivalent to the
self-gravitational energy, ( in Newton's theory -$\frac{3}{5}\left( \frac{%
GM^{2}}{R}\right) $).

The 3rd., stands for gravitomagnetic energy.

An embarrasing problem with (13), is that when R becomes, for instance, R$%
_{0}\leqslant \frac{GM}{2c^{2}}$, we have E 
\mbox{$<$}%
0. This does not violate the positivity of ''total'' energy theorem, but we
interpret as pertaining to a negative energy that represents a repulsive
gravity region ( anti-gravitation ).

This effect could be used to motorize an antigravitational engine. The
technological effort necessary to attain this objective may only succeed at
about 20 years from now. We refer to the above equation for E in the rapidly
spinning case,

\bigskip

E = Mc$^{2}$-$\frac{GM^{2}}{4\varrho }\left[ 1+\left( \frac{a^{2}+\varrho
^{2}}{a\varrho }\right) arctgh\left( \frac{a}{\varrho }\right) \right] $ \ \
\ \ \ \ \ \ \ \ \ \ \ \ \ \ \ \ \ \ \ \ \ \ \ \ \ \ \ \ \ \ \ \ \ \ \ \ \ \
\ \ \ \ \ \ (14).

\bigskip

where $\frac{x^{2}+y^{2}}{\varrho ^{2}+a^{2}}+\frac{z^{2}}{\varrho ^{2}}=1$
\ \ \ \ \ \ \ \ \ \ \ \ \ \ \ \ and \ \ \ $\varrho \geqslant 0$ \ \ \ \ ,
and E stands for the energy inside a surface of constant $\varrho $\ values.

\bigskip

In fact, if we interpret the total energy, as the limit, of E, when the
radial distance goes to infinity, so that the ''entire'' space is involved (
say, put observer at spatial infinity ), the ''positivity'' of \underline{%
total} energy will be preserved, for a positive mass M, while we would still
have, at certain ''radial'' distances $\varrho $ = constant, the possibility
for E 
\mbox{$<$}%
0 ( antigravitation or repulsive gravity ). We must, indeed, not confuse the
''total'' energy ( over all space ), with the energy distribution contents
at radial distance from the source of the field. The ''total'' energy is, of
course, $\lim\limits_{R\rightarrow \infty }E=Mc^{2}>0$. We conclude that
antigravity can be obtained in practice.

\bigskip

\bigskip

{\bf References}

\bigskip

\bigskip \lbrack 1] - Adler, R.; Bazin, M,; Schiffer, M. (1975) -
''Introduction to General Relativity'' - 2nd. edtn., McGraw-Hill, N.Y.

[2] - Virbhadra, K.S. (1990) -Phys. Rev. \underline{D41}, 1086; \underline{%
D42}, 2919. See also paper co-authored by A. Chamorro and J.M.
Aguirregabiria-Gen. Rel. and Grav. 28, (1996) 1393.

\bigskip \lbrack 3] - [12] - See reference [13], page 82, \# 72 to 83.

[13] - Ciufolini, I.; Wheeler, J. A. (1995) - ''Gravitation and Inertia'',
Princeton Univ. Press, Princeton.

[14] - Weinberg, S. (1972) - ''Gravitation and Cosmology'', Wiley, New York.

[15] - Hawking, S. (2002) - ''The Universe in a Nutshell'', Bantam, New York.

[16] - Berman, M.S. (2004) - submitted.

{\bf Acknowledgements}

\bigskip

I recognize enlightening conversations with my intellectual mentors, F.M.
Gomide, and M.M. Som.I am recognized to Prof. James Ipser,whose lectures
helped me to assimilate concepts of vital importance.

\end{document}